\documentclass[aps,letterpaper,twocolumn,showpacs,showkeys]{revtex4}
\usepackage{amsfonts}
\usepackage{amsmath}
\usepackage{amssymb}
\usepackage{graphicx}
\usepackage{textcomp}
\begin{document}
\title{Full Stabilization of a Microresonator based Optical Frequency Comb}
\author{P.~Del'Haye, O.~Arcizet, A.~Schliesser, R.~Holzwarth, T.~J.~Kippenberg}
\email{tjk@mpq.mpg.de}
\affiliation{Max-Planck-Institut f\"{u}r Quantenoptik, 85748 Garching, Germany}


\begin{abstract}
We demonstrate control and stabilization of an optical frequency comb generated by four-wave mixing in a monolithic microresonator with a mode spacing in the microwave regime (86 GHz). The comb parameters (mode spacing and offset frequency) are controlled via the power and the frequency of the pump laser, which constitutes one of the comb modes. Furthermore, generation of a microwave beat note at the comb's mode spacing frequency is presented, enabling direct stabilization to a microwave frequency standard.
\end{abstract}
\maketitle
\revised{}

\textit{Introduction.---}Optical frequency combs
\cite{Holzwarth2000, Diddams2000} have become a powerful tool for
high precision spectroscopy over the past decade and are moreover
used for various applications such as broadband gas sensing
\cite{Thorpe2006}, molecular fingerprinting \cite{Diddams2007},
optical clocks \cite{Diddams2001} and attosecond physics
\cite{Goulielmakis2007}. Frequency comb generation naturally occurs
in mode-locked lasers whose emission spectrum constitutes an
\textquotedblleft optical frequency ruler\textquotedblright\ and
consists of phase coherent modes with frequencies $f_{m}=f_{\text{CEO}}+m\cdot f_{\text{rep}}$ (where $m$ is the
number of the comb mode). Consequently, stabilization of a frequency
comb requires access to two parameters: the spacing of
the modes, which is given by the rate $f_{\text{rep}}$ at which
pulses are emitted, and the offset frequency, given by the carrier envelope
offset frequency $f_{\text{CEO}}$, which can be measured and
stabilized using the powerful technique of self-referencing (by
employing for instance an $f-2f$ interferometer \cite{Jones2000a,Telle1999}). Indeed, these techniques have been
critical to the success of mode locked lasers as sources of optical
frequency combs.

Recently, a \textit{monolithic} frequency comb generator has been demonstrated
for the first time \cite{Del'Haye2007}. This approach is based on continuously pumped fused silica
microresonators on a chip, in which frequency combs are
generated via parametric frequency conversion through four-wave mixing \cite{chang1996},
mediated by the Kerr nonlinearity \cite{Kippenberg2004a,Savchenkov2004a,Ilchenko2006,Agha2007,Maleki2008}. In this energy conserving process, two pump photons are converted into a symmetric pair of sidebands with a spacing
given by the free spectral range (FSR) of the microcavity. This four-wave mixing
process can cascade and give rise to frequency combs spanning up to 500 nm in
the infrared with a mode spacing of up to 1 THz \cite{Del'Haye2007}. The comb modes have been
shown to be equidistant to a fractional frequency uncertainty of one part in
$10^{17}$ relative to the pump frequency \cite{Del'Haye2007}.
\begin{figure}[ptbh]
\begin{center}
\includegraphics[width=\linewidth]{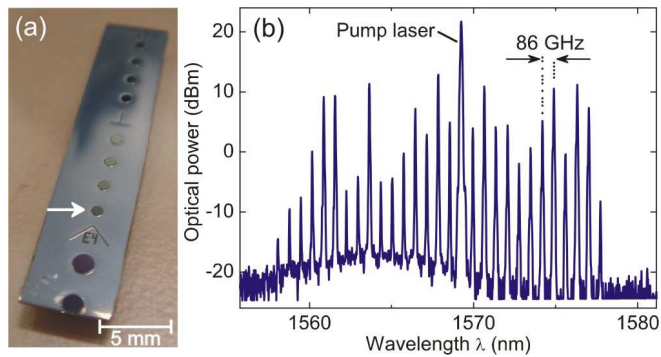}
\end{center}
\caption{(color online). (a) Photograph of a silicon chip with a row
of 750-$\mu$m-diameter monolithic toroidal microresonators made of
fused silica (white arrow). (b) Spectrum of a frequency comb that
has been generated by pumping one of the microresonators with 200 mW
laser power (at 1569 nm), exhibiting a mode spacing in the microwave frequency range (86 GHz).}
\end{figure}

Here we present two major advancements that are neccessary
preconditions for the monolithic comb generator to be viable in
frequency metrology and related applications. First, we demonstrate that it is possible to
control two degrees of freedom of the
microcavity frequency comb (MFC) spectrum, which is required for
full stabilization of the comb spectrum. In contrast to mode locked
lasers, one comb parameter can be \textit{directly} accessed via the
frequency of the pump laser (since it is part of the MFC), whereas
the mode spacing as second degree of freedom is controlled by
changing the optical pathlength of the microcavity via the pump
power dependent refractive index change of the microresonator. This
type of comb control is very robust, since no moveable elements are
involved as in the case of mode-locked femtosecond lasers. The
actuators are used to demonstrate full stabilization of a
microcavity frequency comb. Second and equally important, we show
that smaller mode spacings in the microwave regime can be achieved in monolithic comb
generators. This result thus overcomes a drawback \cite{Cundiff2007} of our previous work
\cite{Del'Haye2007}, which exhibited a mode spacing in the THz range that was not amenable to direct detection with photodiodes. Using larger scale microresonators approaching the mm-range, comb spacings of less than 100 GHz are achieved. 
We show that these combs produce an amplitude modulation in time domain, which is sufficient to directly measure a beat note at the mode spacing frequency. Using this beat note, stabilization of the mode spacing to a microwave frequency standard is demonstrated.

Figure 1 depicts a photograph of this next generation of larger scale
monolithic comb generators with an increased diameter of $D = 750 \ \mu$m
and a corresponding mode spacing of 86 GHz (optical quality factor
$Q\approx 2\cdot10^{7}$, optical linewidth 10 MHz). The measured parametric oscillation threshold of 39 mW agrees well with the theoretical value \cite{Kippenberg2004a} of $P_{\mathrm{th}} \approx 2 \pi^2 \cdot n^2 \cdot \frac{D \cdot A_{\mathrm{eff}}}{\lambda \cdot n_2} \cdot \frac{1}{Q^2} \approx 32 \ \mathrm{mW}$ \footnote{The same threshold bevaviour can be derived by calculating the power that is needed to shift a microcavity resonance by one linewidth via the static Kerr effect.} for an effective mode area of $A_{\mathrm{eff}} \approx \pi \cdot \left( 1.5 \ \mathrm{\mu m} \right)^2$ using $n=1.45$ and $n_2=2.2 \cdot 10^{-20} \frac{\mathrm{m}^2}{W}$ for the linear and nonlinear refractive index of silica. The toroids have been made from microfabricated silica disks with an initial diamameter of 800 $\mu$m, while the reflow process for the generation of surface tension induced toroids has been performed by moving the focus of a CO$_{2}$-laser beam around the circumference of the silica disks. Coupling of laser light into these
resonators is achieved via tapered optical fibers as detailed in prior work \cite{Spillane2003, Cai2000} yielding coupling efficiencies of more than 95\%, an important
prerequisite for high circulating energies within the resonator. Fig. 1(b)
shows a comb spectrum with 86 GHz mode spacing, which is produced by pumping a 750-$\mu$m-diameter microtoroid with 200 mW of continuous wave (cw) power at 1570 nm. The spectral width may be improved by increasing the optical quality factor of the microtoroid ($Q$-factors of up to $10^{9}$ have been attained in millimeter size microspheres \cite{Vernooy1998,BRAGINSKY1989}).


\begin{figure}[ptbh]
\begin{center}
\includegraphics[width=\linewidth]{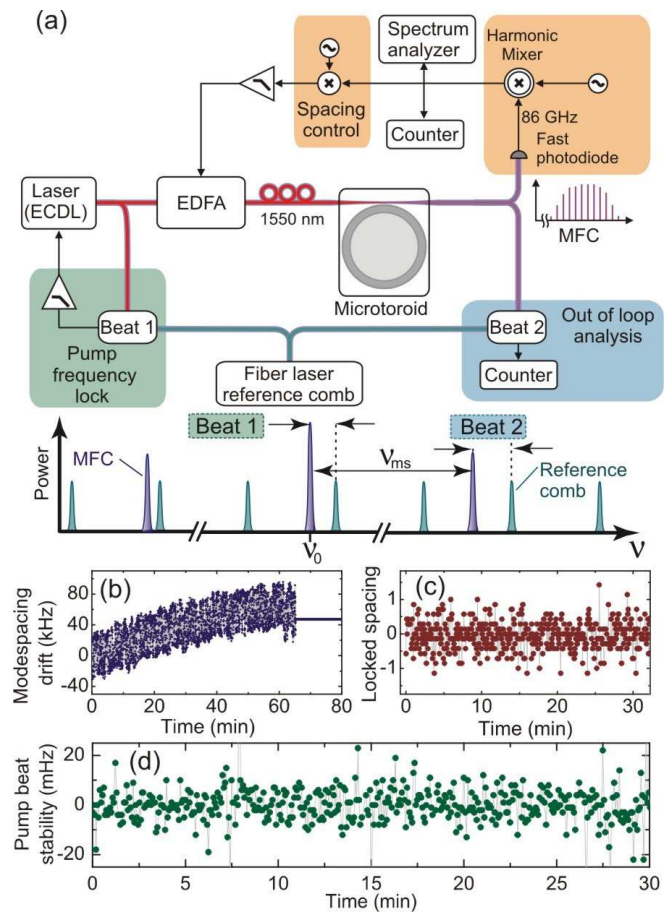}
\end{center}
\caption{(color online). Frequency comb stabilization. (a) Scheme of the
experimental setup used for stabilizing a microcavity frequency comb. The comb spacing is stabilized via the pump power launched into the microresonator, while the pump frequency is stabilized by a phase lock to a reference comb.
Panel (b) displays a measurement of the mode spacing variation of an MFC (mode spacing 400 GHz, measured with ``beat1'' and ``beat2''). The thin line in the right part of the graph has been measured after activating the stabilization of the MFC. Panel (c) and (d) show in-loop counter measurements of the mode spacing and the pump laser frequency of the stabilized microcavity comb at a gate time of 1 s. The standard deviations for the measured data are 400 $\mu$Hz (gate time limited) for the mode spacing lock in (c) and 5.7 mHz for the stabilization of the pump laser in (d).}
\end{figure}

\textit{Stabilization of the microcavity frequency comb.---}The setup to control and stabilize the spectrum of microcavity based frequency combs is depicted in Fig. 2(a). One mode of a toroidal microresonator is pumped with a tunable external cavity diode laser (ECDL) amplified by an erbium doped fiber amplifier (EDFA), leading to the generation of a frequency comb. In contrast to earlier work \cite{Savchenkov2004a,Savchenkov2008} reporting only phase modulation in the four-wave mixing process, we were able to directly measure the 86-GHz mode spacing beat by sending the comb to a fast photodiode (3 dB cut-off at 50 GHz). To measure the 86 GHz signal, it is mixed down to radio frequency (RF) using a harmonic mixer and the 6th harmonic of a local oscillator around 14.3 GHz. The generated RF-signal and thus the comb spacing is stabilized using a phase-locked loop (PLL) that controls the power launched into the microresonator via the diode current of the EDFA. 



The second degree of freedom of the MFC is controlled via the frequency of the pump laser (defining the central mode of the comb), which is phase-locked to an optical reference defined by a mode of a fully stabilized erbium fiber laser based frequency comb with a repetition rate of $\approx$100 MHz \cite{Kubina2005}. Fast control on the pump laser frequency is achieved by actuation of the diode current of the ECDL (actuation bandwidth $\approx$1 MHz). Note that this actuator does not affect the launched pump power, since the subsequent EDFA is operated in saturation, correspondingly amplifying the signal to a constant value. An additional beat note between a MFC mode and a reference comb mode is generated for out-of-loop analysis of the comb stabilization [``Beat 2'' in figure 2(a)]. All local oscillators used for stabilization of the MFC as well as the fiber laser comb are referenced to the same in-house hydrogen maser.

The temporal evolution of the mode spacing without stabilization is depicted in Fig. 2(b), measured with a radio frequency counter at a gate time of 1 s. The first part of the data corresponds to the situation where the active
feedback to the pump power is disabled. A slow mode spacing drift of
approximately 60 kHz/h has been observed, which can be attributed to resonator temperature
drifts during the measurement. Additionally, faster fluctuations with a time
constant of several seconds, which originate from pump power fluctuations and unstable coupling are present. The fluctuations are dramatically reduced when the lock is activated, as can be seen in figure 2(c). Note that the mode spacing in figure 2(b) and 2(c) has been measuring and stabilized using ``beat1'' and ``beat2'' (The experiment has been conducted with a smaller cavity with 400 GHz mode spacing). The recorded beat reveals counter gate time limited fluctuations of less than 1 mHz. Fig. 2(d) depicts the in-loop beat of the pump laser, phase-locked to a reference comb mode. An out-of-loop measurement of the stabilized MFC is presented in the last section.


To obtain a better understanding of the mode spacing of MFCs, a measurement of the Allan deviation $\sigma_A (\tau)$ of both the stabilized and free drifting comb spacing has been conducted [Fig. 3] with a fast photodiode (using the resonator with mode spacing of $\approx$86 GHz). At a gate time of 1 second, the free drifting comb exhibits relative fluctuations of $\sigma_A = 4 \times 10^{-8}$ while the stabilized beat is stable to $\sigma_A = 7 \times 10^{-13}$ relative to the mode spacing frequency. The inset of Fig. 3 shows a spectrum of the stabilized microwave mode spacing beat note, whereby the width of the coherent peak is limited
by the resolution of the spectrum analyzer (10 Hz). 

\begin{figure}[ptbh]
\begin{center}
\includegraphics[width=\linewidth]{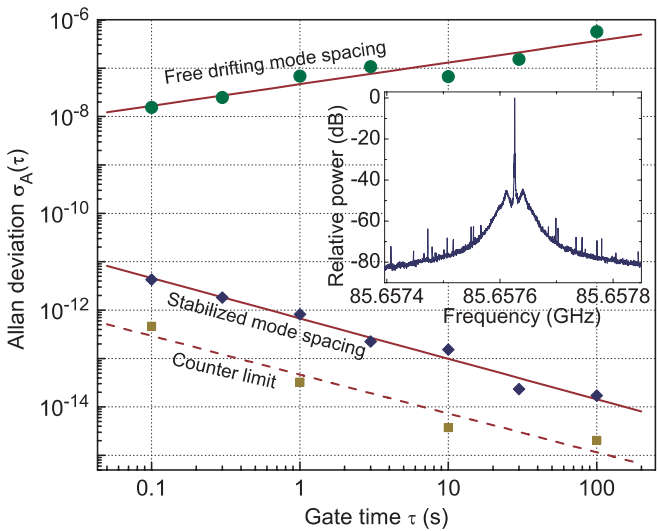}
\end{center}
\caption{(color online). Allan deviation of the free drifting and
the stabilized frequency comb spacing. The inset shows a stabilized
microwave beat note signal of the mode spacing at around 86 GHz with a resolution limited coherent spike.}
\end{figure}



\begin{figure}[ptbh]
\begin{center}
\includegraphics[width=\linewidth]{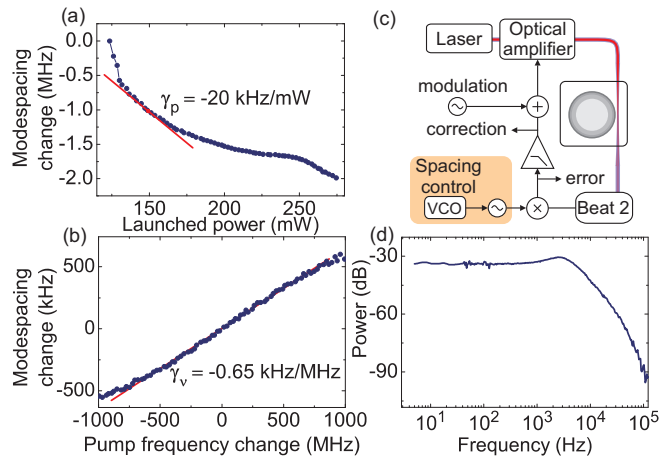}
\end{center}
\caption{(color online). (a) Dependence of the mode
spacing on the power launched into the microcavity. (b) Dependence of the mode spacing on the pump laser frequency (c) Experimental setup for a response measurement of the mode spacing control. (d) Response
measurement of the mode spacing control. The cut-off frequency is around 10
kHz.}
\end{figure}

\textit{Control parameters of the microcavity frequency comb.---}Since the pump laser is part of the frequency comb (similar to the comb generators based on intracavity phase modulators
\cite{KOUROGI1993}), we can describe the comb modes by $\nu_{m}=\nu_\mathrm{0}\pm
m\cdot\nu_{\mathrm{ms}}$ (with the pump frequency $\nu_\mathrm{0}$, the
mode spacing $\nu_{\mathrm{ms}}$ and $m$ being an integer number). To measure
the dependence of the mode spacing $\nu_{\mathrm{ms}}$ on the pump power
$P_{\mathrm{in}}$, one microresonator mode is thermally locked to the pump
laser \cite{Carmon2004} (whereby small fluctuations of the pump detuning are
compensated due to the thermally induced frequency shift of the
microresonator). This thermal self-locking allows for tuning ranges of tens of GHz without losing the resonance of the microcavity \cite{Carmon2004}. After thermally locking the microcavity mode to the pump laser we investigate the influence of pump laser frequency $\nu_\mathrm{0}$ and power $P_\mathrm{in}$ on the comb spacing. Using a matrix notation, the two comb parameters can be written as: 
\begin{equation}
\left( 
\begin{array}{ccc}
\nu_\mathrm{0} \\
\nu_{\mathrm{ms}}-\nu^0_{\mathrm{ms}}   
\end{array} 
\right)	
=
\left( 
\begin{array}{ccc}
1 & 0 \\
\gamma_\nu & \gamma_\mathrm{p}  
\end{array} 
\right)	
\cdot
\left( 
\begin{array}{ccc}
\nu_\mathrm{0} \\
P_\mathrm{in}
\end{array} 
\right)	\ \ \mathrm{,}
\label{eqn:matrixform}
\end{equation}
where $\nu^0_{\mathrm{ms}}$ is the mode spacing at a certain setpoint used for stabilization. The two parameters $\gamma_\mathrm{p}$ and $\gamma_\nu$ describe the influence of the pump laser's power and frequency on the mode spacing, respectively. They are determined by measuring the mode spacing change when varying the power $P_\mathrm{in}$ at a constant frequency $\nu_\mathrm{0}$ and vice versa.
Figure 4(a) and (b) show the result of this measurement yielding $\gamma_\mathrm{p} \approx 20 \ \mathrm{kHz} / \mathrm{mW}$ and $\gamma_\nu \approx 650 \ \mathrm{Hz} / \mathrm{MHz}$ around the chosen setpoint. Thus, having a non-zero $\gamma_\mathrm{p}$ allows for diagonalization of the transfer matrix in equation \ref{eqn:matrixform} and consequently independent control of both comb parameters. 



The main contribution to the mode spacing change with launched power (cf. Fig. 4b) can be explained by the changed temperature of the resonator through absorped optical power. This temperature change leads to a changed refractive index in silica which affects the optical pathlength of the microresonator modes. The thermal contribution to the mode spacing tuning has been quantified by comparing the mode spacing tunability with the pump power dependent frequency of the microcavity mode. The latter has been measured in the same toroid used for stabilization by scanning the laser over a resonance and measuring the maximum detuning frequency with respect to the cold cavity mode at different powers (see reference \cite{Carmon2004} for details on thermal effects in microcavites). The \textit{maximum} resonance shift as a function of pump power has been measured to vary with $\Gamma_\mathrm{p} = -46 \ \mathrm{MHz} / \mathrm{mW}$, which is in good agreement with the measured mode spacing change multiplied by the mode number of $\gamma_\mathrm{p} \cdot m =-20 \ \mathrm{kHz} / \mathrm{mW} \cdot 2244 \approx - 45 \ \mathrm{MHz}/\mathrm{mW}$ (using m = 2244 for a cavity mode at 193 THz and a mode spacing of 86 GHz). Note that for larger pump power tuning ranges the mode spacing dependence becomes more complex, which may be attributed to the generation of additional sidebands contributing differently to the thermal effect of the microcavity and the influence of cross and self-phase modulation \cite{Kippenberg2004a}.

\begin{figure}[ptbh]
\begin{center}
\includegraphics[width=\linewidth]{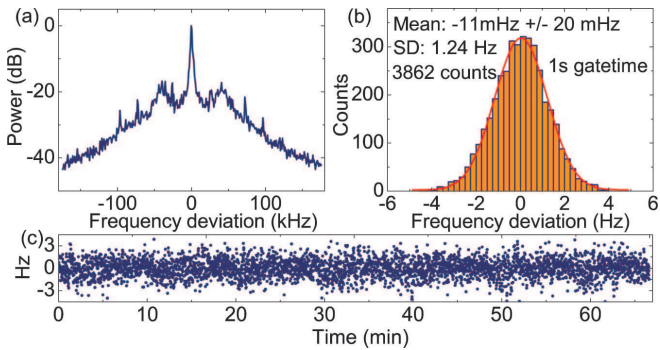}
\end{center}
\caption{(color online). Out-of-loop stability of the microcavity comb (86 GHz mode spacing). (a) \textit{Out-of-loop} beat note of a sideband of the stabilized MFC with a fiber laser reference comb, measured using the setup in figure 2(a). (b) Distribution of the out-of-loop beat note counter measurement data shown below. The deviation from the expected position is smaller than -11 mHz $\pm$ 20 mHz. (c) Counter measurement of the out-of-loop beat note at a gate time of 1 second.}
\end{figure}

Next the locking bandwidth was investigated using the setup in Fig. 4(c). The
control signal for the intracavity power is modulated at varying frequencies
and added to the correction signal of the phase-locked loop, while
simultaneously measuring the correction signal that tries to compensate the perturbation. The result of the
measurement in Fig. 4(d) shows the frequency dependence of the correction
signal, exhibiting a 3 dB cut-off at ca. 10 kHz. This measured value is in good agreement with the expected thermal cut-off
frequency in silica, which can be estimated by $f_{\mathrm{c}}=\frac{\kappa
}{2 \pi \cdot R_{0}^{2}}\approx 16$ kHz \cite{Boyd} (with the thermal diffusivity of silica $\kappa\approx
9\cdot10^{-7}\ \frac{\mathrm{m^{2}}}{\mathrm{s}}$ and $R\approx3\ \mathrm{\mu
}$m being the radius of the toroidal cross section). Thus, the small dimensions of the microcavity makes the thermal response sufficiently fast for the mode spacing stabilization.


\textit{Out-of-loop measurements.---}To quantify the actual stability of the microcavity comb, an additional out-of-loop beat note of a sideband of the stabilized MFC and the reference comb has been generated and measured. The stability of the out-of-loop beat note has been measured with a RF-counter at a gate time of 1 second and shows a standard deviation of 1.24 Hz [Fig. 5].

\textit{Conclusion.---}Stabilization of optical frequency combs
generated by four-wave mixing in on-chip microresonators has been
demonstrated for the first time. It is emphasized that the presented scheme for stabilization does not require \textit{any} moveable parts and is thus highly mechanically robust. Moreover, generation of microwave beat notes is demonstrated allowing locking of the mode spacing to a frequency reference. The stabilization of a microcavity frequency comb in conjuction with mode spacings in the microwave domain is an important step towards a low cost, small sized
frequency comb generator for spectroscopy applications, astrophysical spectrometer
calibration \cite{Murphy2007}, arbitrary optical waveform generation, optical distribution of microwave clock signals and high capacity telecommunication.


We thank T. W. H\"{a}nsch for discussions and suggestions. T.J.K. acknowledges
support via an Independent Max Planck Junior Research Group. This work was
funded as part of a Marie Curie Excellence Grant (RG-UHQ) and the DFG funded
Nanosystems Initiative Munich (NIM). We thank J. Kotthaus for access to clean
room facilities for sample fabrication.


\end{document}